\documentclass{IEEEcsmag}

\usepackage[colorlinks,urlcolor=blue,linkcolor=blue,citecolor=blue]{hyperref}

\usepackage{booktabs}
\usepackage{upmath}

\jvol{XX}
\jnum{XX}
\paper{8}
\jmonth{}
\jname{}
\pubyear{}

\usepackage[colorinlistoftodos,prependcaption]{todonotes}
\presetkeys{todonotes}{inline}{}
\usepackage{hyperref}

\begin{document}
	
\sptitle{ACCEPTED ARTICLE. NOT THE FINAL PUBLISHED ARTICLE.}
\editor{}
 \affil{New York University}

\title{Defending IEEE Software Standards in Federal Criminal Court}

\author{Marc Canellas}

%

\markboth{Canellas -- Defending IEEE Software Standards}{}

\textbf{Copyright Notice:} \textcopyright IEEE 2021. Personal use of this material is permitted. Permission from IEEE must be obtained for all other uses, in any current or future media, including reprinting/republishing this material for advertising or promotional purposes, creating new collective works, for resale or redistribution to servers or lists, or reuse of any copyrighted component of this work in other works. Language per \href{https://journals.ieeeauthorcenter.ieee.org/become-an-ieee-journal-author/publishing-ethics/guidelines-and-policies/post-publication-policies/}{IEEE Post-Publication Policies}.

~
~

\textbf{Final Published Article Citation}: M. C. Canellas, "Defending IEEE Software Standards in Federal Criminal Court," \textit{Computer}, vol. 54, no. 6, pp. 14--23, Jun. 2021. DOI: \href{https://doi.org/10.1109/MC.2020.3038630}{10.1109/MC.2020.3038630}

\begin{abstract}
	~IEEE's 1012 Standard for independent software and hardware verification and validation (IV\&V) is under attack in U.S. federal criminal court. As software spreads through the criminal legal system, scientists, engineers, and IEEE have an essential role in ensuring courts understand and respect IEEE 1012 and IV\&V. If scientists, engineers, and IEEE do not engage, courts will continue to allow unreliable scientific evidence to deprive people of their life and liberty.
	
\end{abstract}

\maketitle

\chapterinitial{D}NA evidence is ``devastating in court'' \cite[para. 38]{Kirchner2017}. Prosecutors and defense attorneys know that DNA evidence all but guarantees a jury's conviction regardless of actual guilt or innocence. Therefore, just the prospect of unfavorable DNA evidence can convince a defendant to plead guilty. But DNA evidence is not an infallible science that only catches the bad guys and exonerates the innocent. Even traditional DNA analysis has caused people to be falsely accused, coerced into false confessions, convicted, and even given the death penalty, because prosecutors and courts did not account for the possibility of erroneous DNA evidence \cite{Murphy2015}. Undeterred, modern DNA analysis through probabilistic genotyping (PG) software is supercharging these catastrophic consequences through trade-secret protected, `internally'-validated black-box technologies.

Traditional DNA analysis uses a one-to-one comparison of directly sampled biological material such as blood or saliva to determine identity and familial relationships. Traditional DNA evidence is so influential in the legal system because it is one of the only forensic science disciplines developed by scientists independent of law enforcement. A landmark report published by the National Research Council (NRC) in 2009 dismissed most forensic evidence as unproven but singled out traditional DNA evidence as the one forensic science worthy of the name \cite{national2009strengthening}. The NRC explained that most other forensic science techniques were ``developed heuristically [meaning] they are based on observation experience, and reasoning without an underlying scientific theory, experiments designed to test the uncertainties and reliability of the method, or sufficient data that are collected and analyzed scientifically'' \cite[p. 128]{national2009strengthening}.

But PG software is nothing like the traditional DNA analysis methods. PG software is the standard bearer of heuristically developed forensic science driven by law enforcement goals not science. PG software uses Markov Chain Monte Carlo methods that purportedly allow for individual identification from tiny samples of DNA that contain a mix of more than one person's genetic material \cite{Kirchner2017}. Where traditional DNA analysis requires high-quality, high-quantity biological materials like blood or saliva, PG software claims to be able analyze low-quality, low-quantity mixtures of DNA like skin cells from objects only touched by the subject of interest \cite{Murphy2015}.

Given the rights and liberty at stake in criminal trials, courts have an obligation to ensure that scientific evidence is reliable. The reliability standard for admitting any scientific evidence is known as the \textit{Daubert} standard in federal court, and as the \textit{Daubert} or \textit{Frye} standard in varying state courts. 

\textit{Daubert} requires that scientific evidence be ``based on sufficient facts or data,'' and ``the product of reliable principles and methods'' which the ``expert has reliably applied... to the facts of the case'' \cite[p. 1]{Tucker2020}. There are many factors that the court should consider, including the theory's testability; the extent to which it has been subjected to peer review and publication; the extent to which a technique is subject to standards controlling the technique's operation; the known or potential rate of error; and, the degree of acceptance within the relevant scientific community \cite{Tucker2020}. Under the \textit{Frye} standard only this last element of general acceptance within the relevant scientific community is necessary for initial admission of scientific evidence.

Defendants have increasingly argued that questions of software reliability under the \textit{Daubert} standard should be assessed via the principles and processes of independent verification and validation (IV\&V) in IEEE Standard 1012-2016, \textit{IEEE Standard for System and Software Verification and Validation} (IEEE 1012) \cite{IEEE_1012}. Therefore, this article focuses exclusively on IEEE 1012. However, other international standards are also relevant, including the \textit{IEEE/ISO/IEC International Standard 29119: Software and systems engineering -- Software Testing} \cite{IEEE_29119}.

Instead of recognizing the shared interests of reliability between law and engineering, the first three federal courts to examine the reliability of STRmix, the most prominent PG software, have each misunderstood and undermined the principles of IV\&V and IEEE 1012. Over the course of three months from November 2019 to January 2020, the federal courts in \textit{United States v. Gissantaner} in Michigan \cite{Gissantaner2019}, \textit{United States v. Tucker} in New York \cite{Tucker2020}, and \textit{United States v. Lewis} in Minnesota \cite{Lewis2020}, each set precedent for undermining these principles which are essential for the safety, security, and economy of our technologically-dependent society.

Scientists, engineers, and IEEE have an obligation to engage with these cases and courts. If we do not, meaningful IV\&V will have no value in criminal litigation, eliminating a necessary process for ensuring that the basis for depriving someone of their life and liberty is reliable and fair.


\section{CONSEQUENCES OF UNRELIABLE SOFTWARE}

In 2011, New York City's Office of Chief Medical Examiner (OCME) deployed the Forensic Statistical Tool (FST), PG software which quickly became a pioneering tool in the criminal legal system \cite{Kirchner2017}.

FST seemed legitimate to many courts. The New York State's DNA Subcommittee certified FST and OCME claimed it could perform beyond the standard FBI DNA practice \cite{Kirchner2017}. At its peak, more than 50 jurisdictions were sending samples to OCME and paying \$1,100 per sample analyzed \cite{Kirchner2017}. 

For years OCME used trade secret protections to keep anyone from seeing FST's source code despite being a public agency. But everything changed in 2016 and 2017 once criminal defendants were able to technically inspect the software and ask questions in court of the OCME analysts operating FST.

It turned out that few, if any, at OCME or New York State's DNA Subcommittee had examined FST in any meaningful way. OCME analysts did not write the code for FST. They hired outside technology consultants \cite{Kirchner2017}. One OCME analyst explained ```We don't know what's going on in that black box'... [E]vidence in older cases should absolutely be retested'' \cite[para. 71]{Kirchner2017}. Another OCME analyst ``conceded that [FST] had not been subject to any external validation or peer review of any kind, and further admitted that there were no documentation or records pertaining to key parts of the internal review process that they claimed to have undertaken'' \cite[p. 99]{Murphy2015}.  As for the DNA Subcommittee which approved FST, one former member testified that they ``met for roughly two hours, three to four times a year,'' ``were not paid for their time,'' and ``did not undertake any independent verification of the assertions made by forensic labs'' \cite[p. 99]{Murphy2015}.

Nathan Adams, the first independent technical analyst allowed to inspect FST concluded in an affidavit that FST was likely not ``developed by an experienced software development team'' and that FST's accuracy ``should be seriously questioned'' \cite[para. 70]{Kirchner2017}. For example, he found that ``a secret function was present in the software, tending to overestimate the likelihood of guilt'' and that the ``actual functioning of the software... did not use the methodology publicly described in sworn testimony and peer-reviewed publications'' \cite[p. 32]{Lacambra2018}.

From a forensic science standpoint, the algorithm did not consider that different people in a mixture could be family and, therefore, share DNA \cite{Kirchner2017}. Dr. Bruce Budowle, an architect of the FBI's national DNA database, testified to the court that the FST's statistical methods were ``not defensible'' \cite[para. 46]{Kirchner2017}. Dr. Budowle also criticized OCME's ``overreliance on `pristine' saliva and samples to test its methods, which do not mirror the ways real crime-scene evidence is degraded by time and weather'' \cite[para. 46]{Kirchner2017}.

Within three months of these revelations, OCME announced it would abandon FST in favor of a newer PG software called STRmix \cite{Kirchner2017}. 

FST had been used for six years in more than 1350 New York criminal trials before it was abandoned in 2017 \cite{Kirchner2017}. But in 2015, two years before these revelations, a lone New York state trial court judge reviewing FST declared that there was ``no scientific consensus in favor'' of FST as a legitimate tool and that defendants ought to be able to appeal their convictions based on FST \cite[p. 17]{Thompson_2019}. But in 2019, two years after FST was abandoned by OCME, that same state trial court judge reaffirmed FST's lack of legitimacy but lamented that his original opinion had become a ``dead letter'' \cite[p. 1]{Thompson_2019}. Incarcerated people's challenges to their FST-based convictions continue to be routinely and summarily denied. Appellate courts consider the use of FST evidence prior to these revelations to not be an abuse of discretion.

For six years, FST evidence was used to charge or convict thousands of Americans despite it being ultimately indefensible, illegitimate, and now, officially and voluntarily abandoned. These people remain in prison, denied even the ability to challenge the legitimacy of their conviction.

IV\&V would certainly have revealed FST's secret function and indefensible statistical methods. Had FST been required to be independently verified and validated prior to deployment, maybe those who were falsely imprisoned in part by FST evidence would be free today. But rather than review and reflect, New York has moved on to STRmix, the focus of the three recent federal court decisions. STRmix has not undergone any more IV\&V than FST, raising the question of whether we are already repeating FST's tragic history.

\section{PROBABILISTIC GENOTYPING SOFTWARE IN CRIMINAL COURT}


PG software, like FST or STRmix, is designed to make identifications from DNA samples that are tiny or contain a mix of more than one person's genetic material \cite{Kirchner2017}. In operation, forensic analysts input their opinion of how many people's DNA are in the sample and what aspects of the DNA to ignore or focus on \cite{Murphy2015}. Then, the PG software uses a combination of forensic DNA techniques, mathematical models, and statistical methods to calculate the likelihood ratio: the likelihood of observing that particular sample assuming the subject of interest was one of the contributors divided by the likelihood of observing that particular sample assuming the subject of interest did not contribute \cite{Murphy2015}.

PG software's reliance on black-box software techniques to analyze low-quality, low-quantity samples often gathered solely from touched objects raises numerous concerns of unreliability because of the variability in how people shed DNA from their skin cells, the risk of contamination and environmental exposure, the variability of forensic lab procedures, the statistical uncertainty in the mathematical models, and the lack of access to the source code due to trade secret protections \cite{Murphy2015}. In particular, PG software developers' staunch opposition to providing any reasonable access to their software is counter to IEEE-USA recommendations supporting the public policy interests of IEEE's over 180,000 U.S. members: ``Intellectual property protections should not be used as a shield to prevent duly limited disclosure of information needed to ascertain whether [systems] meet acceptable standards of effectiveness, fairness, and safety.'' \cite[p. 6]{IEEEUSA_2020}.


Undeterred by these issues, forensic analysts using PG software like FST and STRmix claim they can accurately output a likelihood ratio to the effect of: ``the sample DNA mixture in evidence is 66 trillion times more likely to be observed if the defendant and two unknown others contributed, than if three unknown individuals contributed.'' It is critical to highlight that this is the standard testimony language admitted in court. They do not report any statistical uncertainty suggesting impossibly-certain statistical evidence.

The forensic analysts' use of likelihood ratios indicates they are not focused on following the evidence to answer the ultimate question: ``whose DNA is at the scene?'' Instead, they use the PG software directly guided by the prosecution itself,  asking only ``how confident is the PG software that the person the state intends to convict was at the scene?'' The forensic analysts do not ask whether there may be a different person \textit{more likely} to have contributed, or whether there are ten or a hundred or a thousand other people more likely. By any definition of the phrase, this statistical methodology cannot, alone, prove any defendant's guilt ``beyond a reasonable doubt.'' Yet that is exactly how it is used in court.


Because of these endless issues, biologists, statisticians, and software engineers have implored courts and forensic societies to require that PG software like STRmix follow IV\&V and IEEE 1012 \cite{Adams2018a}. Even the creator of STRmix himself testified in federal court that he ``thought it fair to evaluate STRmix according to the highest---safety critical---IEEE standards'' \cite[p. 1151]{Lewis2020}.

Still, PG software like STRmix still have yet to be independently verified or validated due to resistance from prosecutors, courts, and developers. So, just like FST, it is possible that hundreds or thousands more potential false imprisonments will occur before it is ever truly put through IV\&V. That is, unless scientists, engineers, and IEEE engage.

\section{IEEE 1012: THE STANDARD FOR INDEPENDENT VERIFICATION AND VALIDATION}

Scientists, engineers, and IEEE 1012 have long demanded that safety-critical software and hardware be the right systems built in the right way, and the law should demand this, too. Had the New York courts tried to determine whether FST was the right system built in the right way, New York would not have potentially  1,350 cases of illegal imprisonment. 

Sponsored by the IEEE Computer Society, IEEE 1012 is a universally applicable and broadly accepted process for ensuring that the \textit{right product} is \textit{correctly built} for its \textit{intended use} \cite{IEEE_1012}. IEEE 1012 is used to verify and validate Department of Defense nuclear weapons systems and the National Aeronautics and Space Administration manned space systems and critical space exploration probes, among many other systems.

Verification and validation (V\&V) are interrelated and complementary processes that build quality into any system. Verification is focused on the product itself, providing objective evidence for whether the product conforms to requirements, standards, and practices \cite{IEEE_1012}. Validation is focused on the customer and stakeholders, providing evidence for whether the product is accurate and effective, solves the right problem and satisfies the intended use and user needs in the operational environment \cite{IEEE_1012}. In short, verification ensures that the product is built correctly while validation ensures the right product is built.

Take for example a child's car seat. Successful verification would ensure that the car seat was built to meet the safety regulations. But if the regulations were insufficient to ensure children's safety, verification alone would not be enough. Successful validation would be necessary to ensure the car seat actually kept children safe.

In the context of PG software like FST or STRmix, V\&V would answer the following types of questions \cite{Adams2018a}: Is the model of DNA analysis used by the software the best available, coded as designed, and appropriate for the problem? Does the PG software systematically favor including defendants? How likely are false negatives and false positives? Would outside experts agree with the software's results at each stage of analysis? 

To appropriately perform V\&V, IEEE 1012 requires each software and hardware component be assigned an integrity level that increases depending on the likelihood and consequences of a failure: negligible, marginal, critical (causing ``[m]ajor and permanent injury, partial loss of mission, major system damage, or major financial or social loss'') or catastrophic (causing ``[l]oss of human life, complete mission failure, loss of system security and safety, or extensive financial or social loss'') \cite[p. 196]{IEEE_1012}. As the integrity level increases, so too does the intensity and rigor of the V\&V tasks required by IEEE 1012.

The V\&V process must be independent to avoid potential conflicts of interest that could lead to catastrophic failure. To this end, IEEE 1012 requires technically, managerially, and financially independent V\&V (IV\&V) when testing for any software or hardware where catastrophic consequences could occasionally occur or critical consequences could probably occur \cite{IEEE_1012}. Moreover, letting developers certify their own software is a clear conflict of interest, and the IEEE/ACM Code of Ethics for Software Engineers is clear about the obligation of developers to manage such conflicts \cite{Gotterbarn_1999_Computer}.

\begin{table*}[!t]
	\begin{tabular}{p{6in}}
		Technical Independence: ``[R]equires the V\&V effort to use personnel who are not involved in the development of the system or its elements. The IV\&V effort should formulate its own understanding of the problem and how the proposed system is solving the problem''. ``[T]echnical independence means that the IV\&V effort uses or develops its own set of test and analysis tools separate from the developer's tools.'' And if sharing of tools is necessary, then ``IV\&V conducts qualification tests on tools to assure that the common tools do not contain errors that may mask errors in the system being analyzed and tested.''
		
		\\ Managerial Independence: ``[R]equires that the responsibility for the IV\&V effort be vested in an organization separate from the development and program management organizations. Managerial independence also means that the IV\&V effort independently selects the segments of the software, hardware, and system to analyze and test, chooses the IV\&V techniques, defines the schedule of IV\&V activities, and selects the specific technical issues and problems to act on.'' The IV\&V effort must be ``allowed to submit to program management the IV\&V results, anomalies, and findings without any restrictions (e.g., without requiring prior approval from the development group) or adverse pressures, direct or indirect, from the development group.''
		
		\\ Financial Independence: ``[R]equires that control of the IV\&V budget be vested in an organization independent of the development organization. This independence prevents situations where the IV\&V effort cannot complete its analysis or test or deliver timely results because funds have been diverted or adverse financial pressures or influences have been exerted.''

	\end{tabular} 
	\caption{IEEE Standard 1012-2016 Requirements for Technical, Managerial, and Financial Independent Verification and Validation \cite[p. 198]{IEEE_1012}.} \label{Tab:IVV}
\end{table*}

Full definitions of technical, managerial, and financial independence from IEEE 1012 are provided in Table \ref{Tab:IVV}, but in brief the following must all be separate from the group that oversaw the design and build of the software: the personnel, problem formulation, test and analysis tools for IV\&V (technical); the responsibility for IV\&V (managerial); and, control of the budget for IV\&V (financial)  \cite{IEEE_1012}.

It is essential to remember that the principles of IV\&V are necessary wherever there may be occasional catastrophic or probable critical consequences. IV\&V is a fundamental contributor to the safety and security of modern life, a pillar of safety-critical engineering. Even in operations that do not explicitly adopt IEEE 1012 still uniformly require IV\&V to ensure that system's use is based on objective evidence. For example, the Food and Drug Administration (FDA) requires IV\&V for medical systems \cite{FDA_2002} and the U.S. Nuclear Regulatory Commission (NRC) requires IV\&V for nuclear power plant safety software \cite{NRC_2013}. 


\section{THE IEEE STANDARDS DEVELOPMENT PROCESS}

The primary reason that IEEE 1012 and IV\&V has been so highly regarded throughout the technology world is because they have a proven record for ensuring the reliability of some of the most complex and safety-critical systems in the world, from nuclear power plants and medical systems, to nuclear weapons and space systems. One of the contributors to the worldwide adoption of IEEE 1012 is because it is developed by the IEEE Standards Association (IEEE SA), a world-leading standard setting organization (SSO) with its own reputation for developing reliable and fair standards.

Standards are ``published documents that establish specifications and procedures designed to maximize the reliability of the materials, products, methods, and/or services people use every day'' \cite{IEEESA_2020_Website}. Standards are the basis upon which safety and credibility of new products and new markets are verified, making them fundamental to the modern economy \cite{IEEESA_2020_Website}.




Because standards have such a profound effect on our safety and economy, SSOs like IEEE SA have significant legal obligations with respect to the standards they develop and the processes by which they develop those standards, including contract, intellectual property, and antitrust law \cite{Updegrove2013}. 

Among the many Supreme Court opinions dealing with SSOs, there are two particularly relevant rules SSOs must abide by in order to avoid liability: fair processes and independence. First, standards must be set in a fair manner and courts will look behind the text of standards themselves to ensure that the members are not unfairly skewing the process \cite{Updegrove2013}. Second, SSOs have an obligation to address and eliminate conflicts of interest in the development process. The SSO may be liable even if the SSO is unaware of the action an individual volunteer made and did not approve of nor benefit from the volunteer's action \cite{Updegrove2013}.


%

%

%
%
%
%
%


Heeding the requirements of fair processes and independence, every IEEE SA standard follows a well-defined and documented path from concept to completion, guided by a set of five basic principles and imperatives that ensure fairness and good standards practice during the development cycle \cite{IEEESA_2020_Website}: 

\begin{itemize}
	\item \textbf{Due process}, having highly visible procedures for standards creation and following them;
	\item \textbf{Openness}, ensuring all interested parties can participate actively and are not restricted to a particular type or category of participants;
	\item \textbf{Consensus}, requiring super majority of a voting group to approve a draft of a standard (75\% ballots returned with 75\% of them voting yes);
	\item \textbf{Balance}, ensuring that voting groups include all interested parties and avoid an overwhelming influence by any one party; and,
	\item \textbf{Right of appeal}, allowing anyone to appeal a standards development decision at any point, before or after approval.
\end{itemize}

The IEEE SA principles also adhere to the requirements of the World Trade Organization's (WTO) Decision on Principles for the Development of International Standards, Guides and Recommendations, including transparency, openness, impartiality, and consensus \cite{IEEESA_2020_Website}.

\section{DEFENDING IV\&V AND IEEE 1012}

PG software's use in criminal court is capable of catastrophic failures through false imprisonment and deprivation of people's rights, and therefore, must undergo IV\&V under standards like IEEE 1012. Fundamentally, PG software requires IV\&V by operating in the criminal legal system where its proper and intended use is to cause extensive financial, social, and personal loss.

IEEE-USA emphasizes this requirement for high-risk systems like PG software \cite[p. 5]{IEEEUSA_2020}: 

\begin{quote}
	Before being deployed, high-risk [systems] ought to be independently verified and validated (IV\&V) in accordance with IEEE Standard 1012... and be subject to recurring post-deployment audit, including with respect to their operators. Furthermore, governmental entities should make the reports documenting the required IV\&V and audits of their high-risk [systems] public.
\end{quote}

But the three federal court opinions to address IV\&V and the PG software STRmix pose
a fundamental threat to the legal value of IV\&V and IEEE 1012. The \textit{Tucker} \cite{Tucker2020} and \textit{Lewis} \cite{Lewis2020} courts admitted STRmix evidence without responding to FST's failures, without seriously inquiring about the implications of PG software on people's rights and liberties, and without recognizing the value and track record of IV\&V, IEEE 1012, or IEEE SA. The \textit{Gissantaner} \cite{Gissantaner2019} court rightly precluded STRmix evidence and credited experts' concerns about the software's reliability but claimed without evidence or context that IEEE 1012 is ``not without its faults. For instance, adherence to IEEE actually caused a bug in Excel'' \cite[p. 868]{Gissantaner2019}.

The courts also ignored STRmix's history of now-13 publicly disclosed coding errors that have affected the software's likelihood ratios \cite{STRmix_Miscodes}. None acknowledged that IV\&V would have certainly caught these errors before STRmix had been deployed in thousands of cases, and would catch the errors that will be found as STRmix continues to be used.

The following subsections respond to these three opinions with four clarifications of the principles of IV\&V and IEEE 1012: (1) high-risk software must comply with the IV\&V principles in IEEE 1012, (2) guidance documents are not standards, (3) internal validation is not IV\&V, and (4) reliability requires objective evidence.

\subsection{High-Risk Software Must Comply with the IV\&V Principles within IEEE 1012}

The courts in \textit{Gissantaner} and \textit{Lewis} started their analyses with the explanation that there is no requirement ``that probabilistic genotyping software must comply with the IEEE [1012] standard'' \cite[p. 1151]{Lewis2020} and that ``compliance with IEEE standards is not mandatory, and has not been suggested by any guidance bodies for probabilistic genotyping'' \cite[p. 868]{Gissantaner2019}. The \textit{Gissantaner} court concluded that ``[t]here are no current standards that a lab can be audited against in the forensic community, either in the United States or internationally'' \cite[p. 868]{Gissantaner2019}. Additionally, the courts pointed to the fact that IEEE 1012 itself acknowledges in its ``Notice and Disclaimer'' that ``the existence of an IEEE standard does not imply that there are no other ways to produce, test, measure, purchase, market, or provide other goods and services related to the scope of the IEEE standard'' (\cite[p. 1151]{Lewis2020}, quoting \cite[p. 4]{IEEE_1012}).

The courts' interpretations of IV\&V and IEEE 1012 are gravely mistaken and the recent IEEE-USA position statement made that clear \cite{IEEEUSA_2020}. The courts claim that because IEEE 1012 does not explicitly state ``probabilistic genotyping software must comply with this standard'' that PG software should not have to comply with IV\&V. IEEE 1012 does not explicitly mention nuclear power plants, nuclear weapons, medical systems, or space systems either, but they all rely on the same principles of IV\&V within IEEE 1012 because the principles of IV\&V are universal. As evidenced by its widespread adoption, IEEE 1012 is simply a gold standard version of IV\&V. The courts may as well be saying that if you are cooking food for a informal gathering and thus not explicitly subject to commercial food safety laws, then you have no obligation to ensure that your food does not poison your guests.


IEEE 1012 explains that there are other ways to potentially achieve its ``scope'' of IV\&V because IV\&V is ubiquitous. IEEE 1012 may be the most widely accepted standard process for IV\&V, but just as the FDA and NRC developed their own specific IV\&V processes \cite{FDA_2002,NRC_2013}, it is possible for there to be a separate specific IV\&V process for PG software. However, like all software capable of catastrophic consequences, PG software must follow the fundamental principles of technical, managerial, and financial independence when performing the verification and validation of safety-critical systems. Software engineer codes of ethics \cite{Gotterbarn_1999_Computer} and position statements \cite{IEEEUSA_2020} are clear about this responsibility.

\subsection{Guidance Documents are Not Standards}

The \textit{Lewis} court explained that while STRmix may not comply with IEEE 1012, it does comply ``with all three guidance documents that were specifically adopted for probabilistic systems'' \cite[p. 1151]{Lewis2020}. Most prominent among those guidelines is the one developed by the Scientific Working Group of DNA Analysis Methods (SWGDAM), a group of forensic scientists from international, federal, state and local forensic DNA laboratories \cite{SWGDAM}, laboratories that work for the state, prosecutors, and law enforcement, not the defendants.

SWGDAM and the bylaws that govern the organization violate the basic legal principles that govern SSOs like IEEE SA in the United States. SWGDAM violates due process as there are virtually no procedures publicly available to understand how the guidelines are developed. SWGDAM violates openness and balance as the Chairman of SWGDAM is ``selected and serves at the pleasure of the Director of the FBI's Laboratory Division'' and members are ``appointed by the Chairman based upon recommendations from a Nominating Committee and representation is sought from Federal, State and Local forensic DNA laboratories'' \cite{SWGDAM}. The SWGDAM structure promotes conflicts of interest as the only people contributing to the SWGDAM guidelines for PG systems are those with a direct interest in having evidence from PG software admitted in court.

In sum, it seems that should SWGDAM be treated as an SSO, it would not adhere to the two Supreme Court principles of fair process and independence. Perhaps this is why SWGDAM is careful to label their work product as ``guidelines'' and not ``standards.'' The \textit{Gissantaner} court seemed to acknowledge this distinction when noting that yes, ``[g]uidelines have been issued for the validation of probabilistic genotyping software by the [SWGDAM]; however, they are \textit{merely} guidelines'' (\cite[p. 868]{Gissantaner2019}, emphasis added).

\subsection{Internal Validation is Not IV\&V}

Both the \textit{Tucker} and \textit{Lewis} courts improperly decided that ``internal'' validation was sufficient to show STRmix results were reliable evidence. Letting developers internally validate their own software is a clear conflict of interest that violates the fundamental tenants of IV\&V. Codes of ethics require engineers to disclose and appropriately manage conflicts \cite{Gotterbarn_1999_Computer}. In the case of high-risk software, this means independent, not internal, validation \cite{IEEEUSA_2020}.

In \textit{Tucker}, a New York federal court credited the internal validation by the New York City OCME as sufficient to show reliability, without ever mentioning OCME's dark history with FST \cite{Tucker2020}. In \textit{Lewis}, the court went even further, deeming two internal validation studies as sufficient to conclude that STRmix was reliable: one by the developers of STRmix along with 31 forensic laboratories and a second by an FBI forensic laboratory \cite{Lewis2020}.

The admissions of these internal validation studies were improper. But the \textit{Lewis} court went further by first crediting, then hollowing out  fundamental principles of IV\&V and IEEE 1012.

The \textit{Lewis} court seemed to credit the testimony of Professor Mats Heimdahl, head of the Computer Science and Engineering Center at the University of Minnesota, who explained ``that the purpose of validation is to test and establish the limits of the software and discover when it fails. That is, the persons who validate the software should set out to `break' it, a task that the developers will resist, either consciously or subconsciously'' \cite[p. 1151]{Lewis2020}. Additionally, Dan Krane, Professor of Biological Sciences at Wright State University, explicitly criticized the forensic lab's style of ``internal validation,'' where it paid a third-party to help them interpret their own validation data, characterizing it as an ``almost farcical, ridiculous,'' and unacceptable method of IV\&V \cite[p. 39 of Special Master's Report]{Lewis2020}.

These experts added that this internal validation violated the forensic science recommendation by the President's Council and Advisors on Science and Technology (PCAST) that ``appropriate evaluation of the proposed [forensic DNA] methods should consist of studies by multiple groups, \textit{not associated with the software developers}'' \cite[p. 1147]{Lewis2020}. 

But when making the final decision, the \textit{Lewis} court sought out and found a separate and more permissive passage in the PCAST report to rely on: ``such studies should be performed by or \textit{should include} independent research groups not connected with the developers of the methods and with no stake in the outcome''
(\cite[p. 1147]{Lewis2020}, emphasis in original). 
 The court used this selective license of ``should include'' to agree with STRmix's founder that because there were groups outside the developers ``involved'' in the validation, PCAST's independence requirements were satisfied. The court here decided a question implicating someone's rights and freedoms based on selective choice of sentences in a PCAST report instead of attempting to even superficially understand the principles of IV\&V. 
	
To be absolutely clear, those designing and building safety-critical technology do not look to, accept, or rely on PCAST's imprecise statements of IV\&V. They rely on the standards of technical, financial, and managerial IV\&V like IEEE 1012.

In fact, the \textit{Lewis} court relied on the testimony of STRmix's founder to explain that this complete lack of independence ``makes practical sense'' because ``no forensic laboratory or independent body would have sufficient capability or interest to undertake and publish'' a large validation study \cite[p. 1148]{Lewis2020}. This ignores the reality that there are numerous professional organizations and companies who specialize in IV\&V for software and hardware like STRmix. Surely if independent IV\&V organizations can contribute to the reliability of nuclear power plants, nuclear weapons, medical systems, and space systems, they could do the same for STRmix. Though this would require access to STRmix which has been strictly limited by its developer.

The \textit{Lewis} court acknowledged that the ``reason for insisting that testing be independent is to ensure that the results are reliable and not improperly biased by the developers, either deliberately or subconsciously'' \cite[p. 1148]{Lewis2020}. However, the court then decided that there was no concern here because although the developers of STRmix led the internal validation study, it was a blind study and to the court ``it was unclear how [STRmix developers] could have inappropriately influence the STRmix output, and no questions were raised about the qualifications [of those STRmix developers] who analyzed the data'' \cite[p. 1148]{Lewis2020}. Because there was ``ground truth'' data and ``peer review[ed]'' publications, STRmix must be reliable \cite[p. 1148]{Lewis2020}. 

The court even gave weight to the founder of STRmix's testimony that STRmix `` `very nearly' complies with the IEEE's most stringent safety critical standards''  \cite[p. 1151]{Lewis2020} without requiring the production of any objective supporting evidence and despite the fact that  ``very nearly'' is not a level of compliance for IEEE 1012 or any other standard.

Those claiming the STRmix data was unreliable repeated that these statements did not meet the PCAST requirement that independent research groups have ``no stake in the outcome'' of the validation studies \cite[p. 1147]{Lewis2020}. Clearly, this was a violation of technical, managerial, and financial independence. STRmix developers and the forensic labs have a stake in ensuring that their DNA evidence is considered reliable, in order to shield their product and their work from exclusion.

But while the \textit{Lewis} court agreed that the forensic labs have a ``stake'' in the outcome of the study, the court decided that the internal, ``blind,'' ``peer-review[ed]'' study of STRmix provided ``a reasonable level of assurance of its reliability'' \cite[p. 1148]{Lewis2020}. To the \textit{Lewis} court, a developer-run study was a meaningful approximation of IV\&V.

Ultimately, the \textit{Lewis} court concluded that unless the defendant could establish ``deliberate fraud,'' the STRmix evidence was admissible \cite[pp. 1148-49]{Lewis2020}. Per IEEE 1012 and standard IV\&V practice, the court is placing this burden on the exactly the wrong party.

IEEE 1012's IV\&V tasks include hazard, security, and risk analyses to both qualitatively and quantitatively determine the threats and vulnerabilities to the system from environmental, cybersecurity, and malicious or ignorant actors, among others \cite[pp. 11, 225]{IEEE_1012}. Little could be more na\"{i}ve and inappropriate than for scientists and engineers to presume a high-risk system is immune from threats and vulnerabilities without objective evidence. Therefore, IEEE 1012 requires developers of PG software to have objective evidence answering questions including: Could a forensic scientist manipulate the sample, the data, the software, or the output, without anyone else knowing? If the answer is anything but objective evidence supporting ``no,'' then ``deliberate fraud'' could occur without ever leaving a trace.

\subsection{Reliability Requires Objective Evidence}

A fundamental aspect of IV\&V beyond independence is the requirement of ``objective evidence that the realized system fulfills the requirements, architecture, and design'' \cite[p. 246]{IEEE_1012}. In fact, the only type of evidence accepted for IV\&V is objective evidence.

In \textit{Tucker}, the court relied on the prevalence of peer-reviewed articles, acceptance by forensic labs, and certification by state agencies as evidence of reliability. In theory, these are potentially legitimate sources for determining whether scientific claims are reliable under \textit{Daubert}, but the \textit{Tucker} court shows how these requirements can fail as evidence of reliability in practice.

The question for the court is always about reliability of actual scientific propositions under unbiased, rigorous scientific scrutiny. But to the \textit{Tucker} court, it seemed that developers can simply point to a purported validation study or report with its mere existence alone being the objective evidence.

The \textit{Tucker} court relied on the alleged fact that STRmix and ``its underlying principles have been peer-reviewed in more than 90 articles'' \cite[p. 4]{Tucker2020}. The statement did not explain how only the number of peer review articles supports a finding of reliability, provides any evaluation of the substance of those articles, or shows its general acceptance by communities outside of narrow forensic science. The court did not explain how this positive inference would be affected by the fact that many of those articles were published by developers of STRmix or by forensic laboratories who rely on STRmix evidence in court, almost exclusively in narrow forensic science journals. To achieve true general acceptance for allegedly novel statistical software methods, it would be critical to publish in peer-reviewed journals broadly accepted as authorities in statistics, software engineering, and computer science -- many operated by IEEE and their members.

In truth, the language in the court's decision suggests the court did not even know first-hand whether there existed 90 peer-review articles validating STRmix. The court cited a separate court opinion from three years earlier for that number, \textit{United States v. Pettway} \cite[p. 4]{Tucker2020}. The \textit{Pettway} court also did not review 90 articles but was instead quoting a single statement by the director of a New York state forensic lab who only claimed that there were 90 articles based on conversations with the STRmix developers.

So, it seems the \textit{Tucker} court admitted software evidence used to deprive someone of their liberty based in part on the developer's mere claim that there existed 90 peer-reviewed articles of the software. Articles which were never produced nor reviewed by the court for their reasoning, legitimacy, conflicts of interest, journal authority, or experimental results, if any.

The \textit{Tucker} court also relied on the fact that ``[k]nowledgeable bodies have evaluated the software and approved its use, such as the DNA Subcommittee of the New York State Commission on Forensic Science'' \cite[p. 4]{Tucker2020}. The court again did not explain how the DNA Subcommittee certifies software and did not mention the problematic history of the DNA Subcommittee in New York. Even a cursory glance would have found the problems with FST described above and the public knowledge that the DNA Subcommittee ``did not undertake any independent verification of the assertions made by forensic labs'' \cite[p. 99]{Murphy2015}.

Lastly, the \textit{Tucker} court explained that ``STRmix is currently in use in over forty states and federal laboratories in the United States and in at least thirteen other countries.'' Here, the \textit{Tucker} court cites another opinion, \textit{United States v. Christensen} \cite[p. 4]{Tucker2020}, without revealing that this alleged fact from \textit{Christensen} is actually just oral testimony from an FBI forensic analyst. Moreover, the \textit{Tucker} court did not elaborate on whether the certification methods in those forty states or thirteen other countries were relevant to the principles of IV\&V or the legal standard of  \textit{Daubert} reliability. This assumption that if other courts accept it, it must be reliable, makes no attempt to respect the troubled history of forensic science. For decades courts allowed people to be imprisoned and put to death solely based on now-discredited forensic science techniques like hair and bitemark identification \cite{national2009strengthening}.

Again and again, mere claims of evidence were taken as objective evidence of reliability without any sense of investigation by the judge: the sole person responsible for ensuring foundational reliability. The federal rules of \textit{Daubert} demand that the judge look at the reliability of actual scientific propositions. But the \textit{Tucker} court only relied on verbal testimony and single-sentence quotes from reports when executing its responsibility to ensure fairness and justice in one of the most consequential acts of our legal system.

\section{OUR RESPONSIBILITIES}

In a world increasingly dominated by technology, these three federal criminal court opinions, \textit{Gissantaner}, \textit{Lewis}, and \textit{Tucker}, describe an American legal system that cannot appropriately define or value IV\&V. Precedence will soon codify these decisions undermining the principles of IV\&V present in IEEE 1012. This is call to arms. Individually and collectively, we scientists and engineers have meaningful work before us.

Whatever one's views are of the criminal legal system, technology is becoming an integrated part of modern law enforcement and government prosecution. Scientists and engineers have a obligation to make sure that the technologies informing the deprivation of life and liberty work reliably in a way that promotes human values and ensures trustworthiness \cite{Gotterbarn_1999_Computer,IEEEUSA_2020}. In this vein, peer-review is not only a tool for those undermining IV\&V, it can be tool for promoting the application of IV\&V to these systems. Any serious examination of forensic technologies will be read with great interest by attorneys and, hopefully, the courts.

Scientists and engineers can also join with those trying to bring accountability to these systems either through developing technical standards with the IEEE SA or by getting involved in the litigation themselves. Legal organizations and the courts, as evidenced by these opinions, are constantly searching for technical experts to help them determine whether a newly-deployed technology is reliable.

IEEE as an organization has a clear mandate as well. IEEE should:

\begin{itemize}
	\item File amicus briefs which are third-party legal documents advising courts of the relevant additional information the court ought to consider in a particular case, including the broader implications of their decisions. This article can be viewed as an outline for such an amicus brief.
	\item Develop IEEE SA standards explicitly for forensic software and hardware with catastrophic and critical consequences like PG software and face recognition systems \cite{IEEEUSA_2020}.
	\item Publicly comment on other standards organizations forensic software standards.
	\item Clarify that the description of independence in Standard 1012 accounts for conflict of interests by excluding those with any particular stake in the outcome. IV\&V clearly excludes V\&V by forensic labs who, though not financially dependent on the developers, have a shared interest in the software's acceptance.
\end{itemize}

In reflection our responsibilities, I leave you with the prescient words of Norbert Weiner, the father of cybernetics: 

\begin{quotation}
	``Whether we entrust our decisions to machines of metal, or to those machines of flesh and blood which are bureaus and vast laboratories and armies and corporations, we shall never receive the right answers to our questions unless we ask the right questions. ... The hour is very late, and the choice of good and evil knocks at our door'' \cite[pp. 185-86]{Weiner1989}.
\end{quotation}

\bibliographystyle{IEEEtran}
\bibliography{IEEE_IVV_Comp}

\begin{IEEEbiography}{Marc Canellas} is a third-year law student at the New York University School of Law and the Vice-Chair of the IEEE-USA AI Policy Committee. Dr. Canellas received his Ph.D. in aerospace and cognitive engineering from the Georgia Institute of Technology. His professional work and research interests include public defense and the governance of human-machine systems, particularly with respect to carceral technology and product liability. Contact him at marc.c.canellas@gmail.com. The contents of this article are solely the responsibility of the author and do not represent IEEE.
\end{IEEEbiography}

%

\end{document}